\documentclass[aps,pra,twocolumn,superscriptaddress,byrevtex,showpacs,preprintnumbers,amsmath,amssymb,floatfixm]{revtex4}
\pdfoutput=1
\usepackage{wasysym}
\usepackage{graphicx}
\usepackage{dcolumn}
\usepackage{bm}
\usepackage{color}

\begin{document}
\title{ac electric trapping of neutral atoms}

\author{Sophie Schlunk}
\email{schlunk@fhi-berlin.mpg.de}
\affiliation{Fritz-Haber-Institut der Max-Planck-Gesellschaft,
Faradayweg 4-6,
  14195 Berlin, Germany}
\affiliation{FOM-Institute for Plasmaphysics Rijnhuizen, P.O. Box
1207, 3430 BE Nieuwegein, The Netherlands}
\author{Adela Marian}
\author{Wieland Sch\"ollkopf}
\author{Gerard Meijer}
\affiliation{Fritz-Haber-Institut der Max-Planck-Gesellschaft,
 Faradayweg 4-6, 14195 Berlin, Germany}
\date{\today}
\begin{abstract}
We study the dynamic behavior of ultracold neutral atoms in a
macroscopic ac electric trap. Confinement in such a trap is
achieved by switching between two saddle-point configurations of
the electric field. The gradual formation of a stably trapped
cloud is observed and the trap performance is studied versus the
switching frequency and the symmetry of the switching cycle.
Additionally, the electric field in the trap is mapped out by
imaging the atom cloud while the fields are still on. Finally, the
phase-space acceptance of the trap is probed by introducing a
modified switching cycle. The experimental results are reproduced
using full three-dimensional trajectory calculations.
\end{abstract}


\maketitle

\section{Introduction}

The trapping of neutral particles has paved the way for remarkable
achievements in atomic and molecular physics, culminating with the
observation of degenerate Bose and Fermi gases \cite{Anderson95,
DeMarco99}. The ability to confine the ground state of neutral
particles is becoming important for an increasing number of
experiments, in particular for collision studies \cite{Lara06}.
The ground state of any neutral particle is always attracted
towards a point of maximum field, but it cannot be trapped in a
static field, as static fields cannot possess a maximum in free
space \cite{Wing84, Ketterle92}. Trapping in the ground state is
therefore only possible when electrodynamic fields are used. For
example, any sublevel of the ground state can be confined in the
laser field maximum of a dipole trap \cite{Grimm00}, which has
been successfully demonstrated for both atoms \cite{Chu86} and
molecules \cite{Takekoshi}. Additionally, paramagnetic particles
can be trapped using ac magnetic fields, as shown for ground-state
Cs atoms \cite{Cornell91}.

Trapping by ac electric fields is a more versatile method,
applicable to any ground-state atom or molecule. For polar
molecules strong confinement can be achieved via the first-order
Stark interaction. For atoms and nonpolar molecules, trapping is
based on the second-order Stark interaction between an external
electric field and the induced dipole moment. Similarly to
trapping of ions in a Paul trap \cite{Paul90}, three-dimensional
confinement in an ac electric trap is achieved by alternating
between two saddle-point configurations of the electric field. The
first configuration has attractive (focusing) forces along one
direction and repulsive (defocusing) forces along the other two
directions, while in the second configuration the roles of the
forces are reversed. Dynamic confinement of the particles is
obtained by switching between these two configurations at the
appropriate frequency.

The first demonstration of ac electric trapping was carried out
for ammonia molecules using a cylindrically symmetric trap with a
depth of several millikelvins \cite{Jacqueline05}. Similar trap
depths were later obtained for the same molecule with a linear ac
trap \cite{Schnell07}. In the meantime, Katori and coworkers
achieved trapping of about 100 ground-state Sr atoms with a
lifetime of 80 ms in a microstructured ac trap on a chip
\cite{Katori06}. Recently, we demonstrated trapping of about
10$^5$ Rb atoms in a 1 mm$^3$ large and several microkelvins deep
trap, with a lifetime of about 5 s \cite{Schlunk2007}. Our trap is
also based on the cylindrically symmetric geometry suggested by
Peik \cite{Peik99}. It allowed for the first direct visualization
of the dynamic confinement in an ac trap using absorption images
taken at different phases of the ac switching cycle. Subsequently,
trapping of ultracold Rb atoms was also obtained with a
three-phase ac electric trap \cite{Pinkse07}, as proposed by
Shimizu and Morinaga \cite{Shimizu92, Morinaga94}. A recent paper
describes in detail the ac trap geometries currently used to
electrically confine neutral atoms and molecules
\cite{Jacqueline06b}.

In this paper, we present a detailed experimental investigation of
the dynamics in the ac trap and study the dependence of the atom
number on the switching-cycle parameters. Compared to our previous
measurements \cite{Schlunk2007}, we have optimized the number of
trapped atoms by introducing an evaporative cooling stage before
loading the atoms into the ac trap. As the depth of the ac trap is
small, the reduction in temperature helped increase the number of
confined particles, which in turn improved the quality of our
images. The experimental data are compared throughout the paper
with results of numerical simulations based on classical
trajectory calculations.

This paper is organized as follows. We start by describing the
experimental sequence used to load the atoms into the ac trap in
Sec.~\ref{sec:ExperProcedure}. The ac trap is presented in detail
in Sec.~\ref{sec:Design} and the theory model we use for the
simulations is also briefly discussed. Next, we show how the atoms
can be used to probe the electric fields in the trap in
Sec.~\ref{sec:InTrap}. The formation of a trapped cloud in the ac
trap is studied in Sec.~\ref{sec:CloudEvolution} by imaging the
atoms after a gradually increasing number of switching cycles.
This also leads to a measurement of the ac trap lifetime. Then, in
Sec.~\ref{sec:Performance}, the dynamics of the atoms in the ac
trap is visualized at different phases of the ac switching cycle.
Additionally, the dependence of the trapped atom number on the
trapping frequency and the symmetry of the switching cycle is
measured. Furthermore, the mean kinetic energy of the atoms is
determined at different phases of the switching cycle. In
Sec.~\ref{sec:PhaseJump} we study the phase-space acceptance of
the ac trap by introducing a sudden change in the switching cycle.
Finally, we summarize our results in Sec.~\ref{sec:conclusions}.

\section{Experimental procedure}\label{sec:ExperProcedure}

In the experiment, the $^{87}$Rb atoms are first collected in a
six-beam magneto-optical trap (MOT) loaded from a Zeeman slower.
After a short compression of the MOT, optical molasses cooling,
and optical pumping, the atoms are transferred into a spatially
overlapped quadrupole magnetic trap. About $5 \times 10^8$ atoms
in the $F = 2$, $m_F = 2$ hyperfine sublevel at a temperature of
600 $\mu$K are trapped in the magnetic trap which is characterized
by a field gradient of 270~G/cm along its symmetry axis. A short
evaporative cooling stage is applied next by linearly ramping a
radio frequency from 35 to 6~MHz in 8.3~s. At the end of the
evaporative cooling stage, the field gradient is reduced to 65
G/cm. Approximately $2 \times 10^7$ atoms remain magnetically
trapped at a temperature of 30 $\mu$K. The quadrupole magnet is
then moved horizontally in 3~s to a second quartz cell located
42~cm away \cite{Lewandowski03}. This second ultrahigh-vacuum
chamber houses the ac trap. The transfer is carried out using a
precision translation stage  allowing for accurate overlap of the
cloud with the center of the ac trap. Altogether, the cooling,
trapping, and transport of the atoms take about 25 s. At the final
position, the magnetic field is switched off and, once it has
completely disappeared, high voltage is applied to the ac trap
electrodes. After a variable trapping time, the trap electrodes
are switched back to ground and the confined atoms are detected by
absorption imaging. In a typical experiment, the atoms are imaged
0.1 ms after the high voltage has been turned off, thereby
reflecting the spatial distribution at the time of switch-off. The
number of atoms is determined with an accuracy of about 5\% using
a two-dimensional Gaussian fitting procedure.

\section{Trap design}\label{sec:Design}

\begin{figure}[pt]
\includegraphics[scale=0.58]{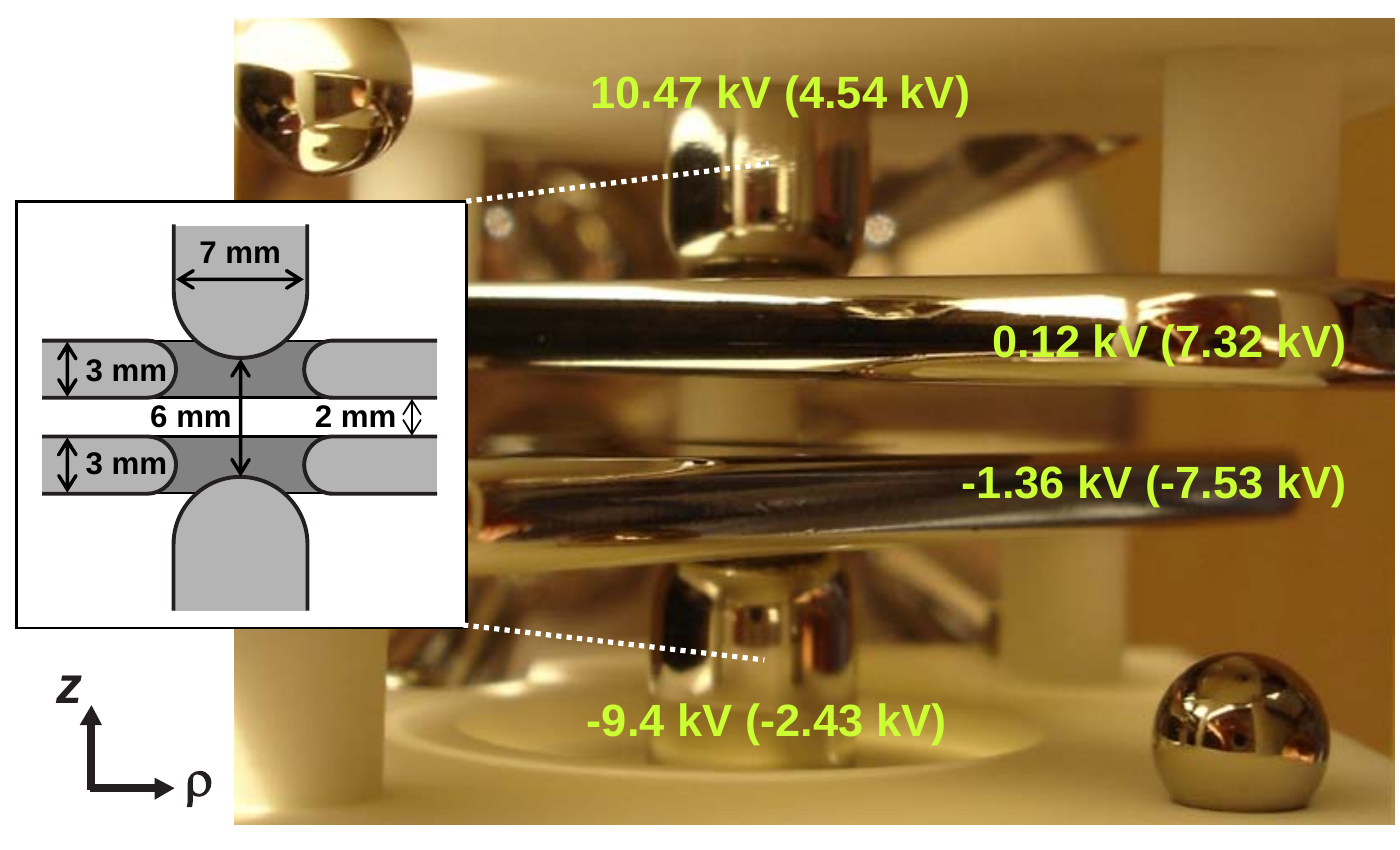}
\caption{Photograph of the ac trap. The inset on the left shows a
schematic of the trap where the dimensions of the electrodes and
the distances between them are indicated. The voltages applied to
the electrodes during the $\rho$-focusing ($z$-focusing) phase are
written on each electrode.}
\label{fig:PictureTrap}
\end{figure}

In the following, we give a description of the ac trap and briefly
introduce our model for theory simulations.

The ac trap consists of two end cap electrodes and two ring
electrodes as shown in Fig.~\ref{fig:PictureTrap}. The inset is a
schematic cross section of the cylindrically symmetric trap, while
the photograph zooms in on the trap used in the experiment. The
four electrodes are made of nonmagnetic stainless steel and are
mounted between two macor plates using macor spacers. The two ring
electrodes are located between the end caps, but the ring opening
is not visible in the photograph because the picture is taken from
the side. As indicated in the schematic, the end caps have a
hemispherical shape with a diameter of 7 mm and are separated by 6
mm. The ring electrodes are 3 mm thick, corresponding to an inner
semicircular shape with a radius of 1.5 mm. They have an opening
diameter of 6.7 mm and are separated by a 2 mm gap. All electrodes
were highly polished before the trap was assembled and were
subsequently conditioned to withstand increasingly high voltages.

A superposition of a static dipole field and an alternating
hexapole field is used to switch between the two saddle-point
configurations \cite{Peik99}. Because only the hexapole field is
switched, the field strength remains constant at the center of the
trap. Additionally, a dc quadrupole field is applied to counteract
gravity. The two saddle-point configurations are referred to as
$\rho$ and $z$ focusing. To alternate between these two
configurations, two different voltage sets are applied to the four
electrodes, as indicated in the photograph in
Fig.~\ref{fig:PictureTrap}. The ac switching frequency is given by
$1/T$, where $T$ is the sum of the durations of the
$\rho$-focusing and $z$-focusing phases, i.e., the duration of one
full switching cycle.

\begin{figure}[t]
\includegraphics[scale=0.45]{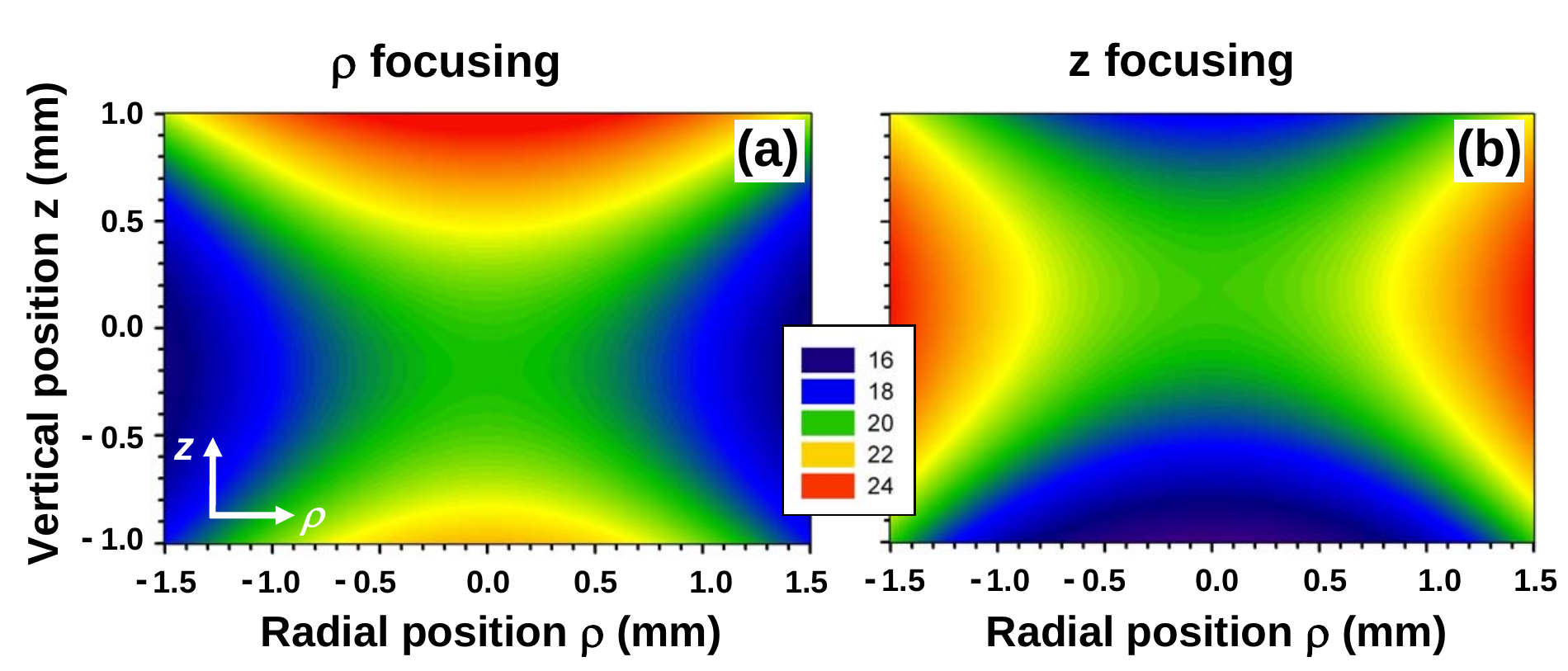}
\caption{Calculated electric field (in kV/cm) in the trap for the
$\rho$-focusing (a) and  $z$-focusing (b) configurations.}
\label{fig:TrapFields}
\end{figure}

The calculated fields that the atoms experience in the two
different switching configurations are plotted in
Figs.~\ref{fig:TrapFields}(a) and \ref{fig:TrapFields}(b). For the
$\rho$-focusing configuration in Fig.~\ref{fig:TrapFields}(a), the
field has a maximum in $\rho$ and a minimum in $z$, and the
highest field value is found at the top of the picture. For $z$
focusing the situation is reversed; the field has the maximum in
$z$ and the highest fields are positioned symmetrically on the
left and right sides of the picture. Note that the saddle points
are displaced in $z$ due to the dc quadrupole field used for
gravity compensation. The $z$-direction gradients for this field
have the same value for both trapping configurations.

To describe the fields in the trap, we approximate the electric
potential by a multipole expansion. Taking the cylindrical
symmetry into account, the first five terms read
\begin{eqnarray}
\Phi(\rho, z)= \Phi_0 + \Phi_1 \frac{z}{z_0} + \Phi_2 \frac{z^2 -
\rho^2/2}{z_0^2} + \Phi_3 \frac{z^3 - \frac{3}{2} z \rho^2}{z_0^3} + \nonumber \\
\Phi_4 \frac{z^4 - 3 \rho^2 z^2 + \frac{3}{8} \rho^4}{z_0^4} +
\Phi_5 \frac{z^5 - 5 z^3 \rho^2 + \frac{15}{8} z \rho^4}{z_0^5},
\label{eq:MultipoleExpansion}
\end{eqnarray}
where $2 z_0 = 6$ mm is the distance between the end cap
electrodes and the $\Phi_i$ characterize the different multipole
terms. The first term describes a constant potential, the second
term represents a dipole field, the $\Phi_2$ term is the
quadrupole potential, and the $\Phi_3$ term is the hexapole
component. The $\Phi_4$ and $\Phi_5$ terms are octupolar and
decapolar field components. These last two higher-order terms
represent undesirable nonlinearities in the system.

Simulations are carried out to model the experimental results.
First, the electric fields for our trap geometry are calculated
using a commercial finite-element program (COMSOL). A multipole
series up to the fifth term as shown in
Eq.~(\ref{eq:MultipoleExpansion}) is then fitted to these fields.
For the trap geometry presented in Fig.~\ref{fig:PictureTrap}, the
multipole term components are $\Phi_0\!=\!-285$~V,
$\Phi_1\!=\!5992$~V, $\Phi_2\!=\!604$~V, $\Phi_3\!=\!3265$~V,
$\Phi_4\!=\!222$~V, and $\Phi_5\!=\!737$~V for $\rho$ focusing,
and $\Phi_0\!=\!232$~V, $\Phi_1\!=\!5991$~V, $\Phi_2\!=\!607$~V,
$\Phi_3\!=\!-3263$~V, $\Phi_4\!=\!223$~V, and $\Phi_5\!=\!687$~V
for $z$ focusing. The forces acting on the atoms are then derived
and trajectory calculations are carried out by numerically
integrating the equations of motion. A fine grid in phase space is
used to simulate the initial distribution of atoms. Throughout the
paper, simulation results will be presented in conjunction with
experimental data.

\section{Mapping of the electric fields}\label{sec:InTrap}

\begin{figure}[pt]
\includegraphics[scale=0.45]{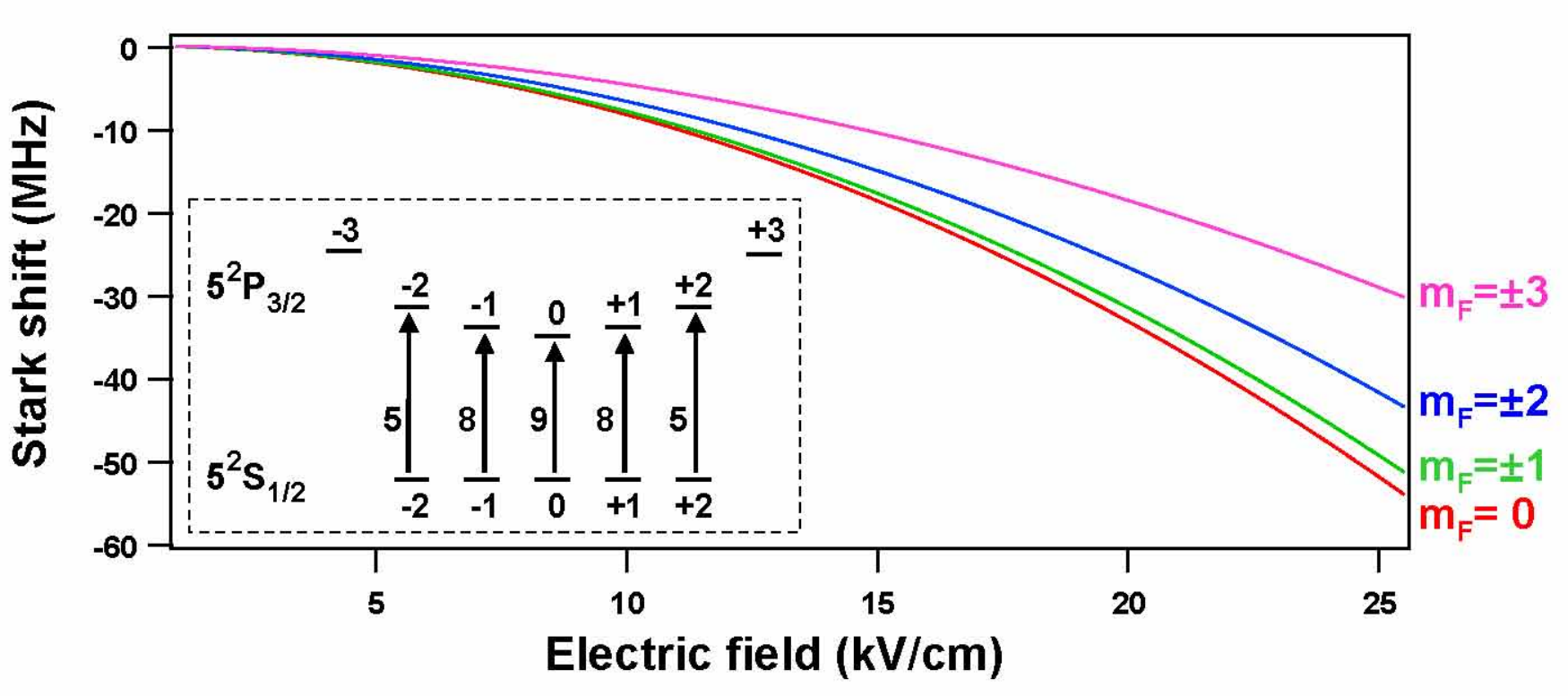}
\caption{Stark shift of the $5\,^{2}S_{1/2}(F=2) \rightarrow
\,5\,^{2}P_{3/2}(F=3)$ transition used for probing. For the upper
state the Stark shift is different for the $|m_F|$ sublevels,
while for the lower state all $m_F$ sublevels are degenerate. This
can also be seen in the inset which shows the probing transition.
All degenerate sublevels of the $5\,^{2}S_{1/2}(F=2)$ state are
populated in the trap; the linearly polarized probe beam drives
only transitions with $\Delta m_F = 0$. The relative transition
strengths are indicated next to the arrows.}
\label{fig:StarkShift}
\end{figure}

We will now describe how the atoms can be used to sensitively
probe the actual electric fields in the trap. This is important
because small inaccuracies and misalignments of the electrodes can
have a strong impact on the trapping fields.

To probe the electric fields in the ac trap, the usual
experimental sequence is followed with the only exception that the
electric fields are kept on while taking absorption images of the
cloud. The atoms are confined in the ac trap for a short trapping
time during the first switching cycle and the electrodes are
switched to ground after obtaining the absorption images. To
compensate for the resulting Stark shift, the absorption beam is
detuned by several megahertz while the atoms are probed. We have
already seen in Fig.~\ref{fig:TrapFields} that the electric field
strength varies significantly over the size of the atom cloud.
Therefore, the position of an atom determines whether it is on
resonance with the probing transition. Note that the absorption
beam crosses the entire cloud and due to the cylindrical symmetry
it can be thought of as passing once from left to right through
the fields shown in the contour plots in
Fig.~\ref{fig:TrapFields}.

In the electronic $5\,^{2}S_{1/2}(F=2)$ ground state, the $m_F$
sublevels remain degenerate in an electric field. In the
$5\,^{2}P_{3/2}(F=3)$ excited state, however, the degeneracy is
lifted due to the tensor polarizability \cite{Krenn97}, as seen in
the inset of Fig.~\ref{fig:StarkShift}. This gives rise to four
possible energy differences (Stark shifts) between the two states,
which are plotted in Fig.~\ref{fig:StarkShift}.

\begin{figure}[pt]
\includegraphics[scale=0.45]{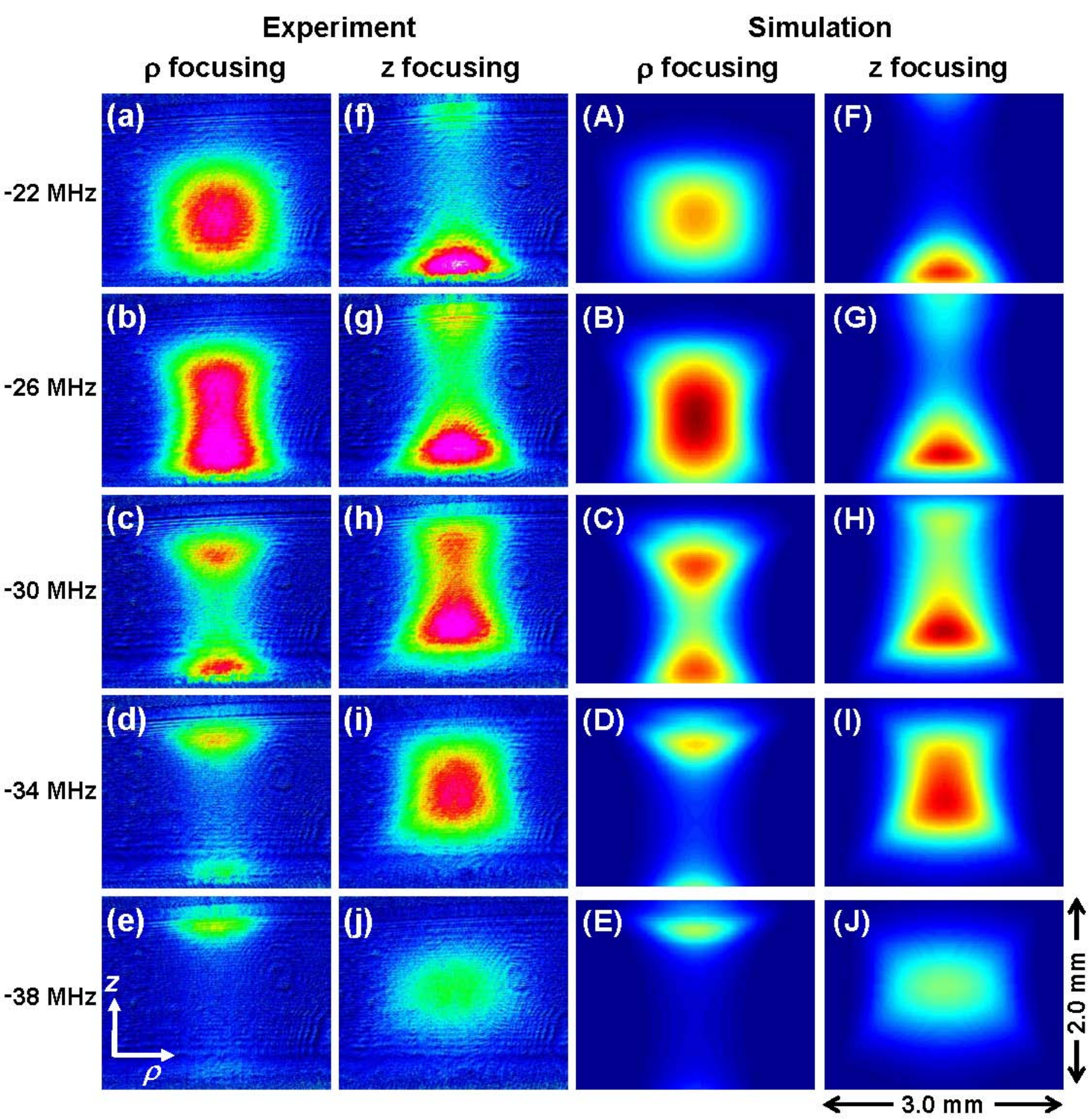}
\caption{Absorption images (a)$-$(j) and simulations (A)$-$(J) of
the atom cloud for different probe detunings with the electric
fields still on. The first column displays images recorded during
$\rho$ focusing, when the atoms are kept in the trap for 9.83 ms.
For the second column, the atoms are probed in the $z$-focusing
configuration after a trapping time of 9.84 ms. All absorption
images are averaged five times. The pictures in the third and
fourth columns show the corresponding simulations.}
\label{fig:InTrap}
\end{figure}

In Fig.~\ref{fig:InTrap} the first two columns show images of the
atoms in the $\rho$-focusing and $z$-focusing configurations,
respectively. The third and fourth columns display simulation
results. For the $\rho$-focusing pictures in the first column, the
atoms are trapped for 9.83 ms, just before switching to $z$
focusing. For the measurements shown in the second column, the
atoms are imaged immediately after switching to the $z$-focusing
configuration, namely, after a trapping time of 9.84 ms. The
difference in trapping times for the two configurations is tiny
and ensures that the atom density distributions are identical.
Additionally, for these short trapping times the atom cloud is
dense, resulting in a good absorption signal. The experiment is
repeated, varying the probe detuning from -22 to -38~MHz.
Depending on the detuning, different electric field regions are
visible, thus enabling direct probing of the field distributions
shown in Fig.~\ref{fig:TrapFields}. For instance, the atom
distributions in Figs.~\ref{fig:InTrap}(c) and \ref{fig:InTrap}(g)
look like an hourglass, thereby resembling the shapes of the
fields in Fig.~\ref{fig:TrapFields}, with the saddle points
located at the waist of the hourglass. As expected, the waists are
displaced in $z$ for the two pictures. The simulations, labelled
with matching capital letters, agree well with the measurements.

Looking at the pictures in the first column, the lower fields
corresponding to a small probe detuning are found in the center
[Fig.~\ref{fig:InTrap}(a)]. The higher fields at the top and at
the bottom become visible for the larger detunings of -34 MHz in
Fig.~\ref{fig:InTrap}(d) and -38 MHz in Fig.~\ref{fig:InTrap}(e).
Note that the field strength is higher at the top than at the
bottom, in agreement with Fig.~\ref{fig:TrapFields}(a). The trend
is reversed for the $z$-focusing configuration shown in the second
column. For a small detuning of -22 MHz [Fig.~\ref{fig:InTrap}(f)]
atoms at the top and at the bottom are on resonance. With
increasing detuning only atoms at the center of the trap are
visible, confirming that along the $z$ axis the highest fields are
in the center of the trap as seen in Fig.~\ref{fig:TrapFields}(b).
Additionally, in the case of $z$ focusing, the higher fields
towards larger $\rho$ play a role. Nonetheless, the highest fields
on the left and on the right side of the trap are not visible
because the atom cloud is more dilute at larger $\rho$, and
therefore the atom density is too low to image these trap regions.

As mentioned in Sec.~\ref{sec:Design}, our simulations are carried
out using a finite-element program to determine the electric field
in the trap. The probe beam attenuation by an atom cloud with a
Gaussian density distribution is calculated for the simulated
electric fields. The resulting theory plots are shown in
Figs.~\ref{fig:InTrap}(A)$-$\ref{fig:InTrap}(J). In the magnetic
trap only the $m_F=+2$ sublevel of the $5\,^{2}S_{1/2}(F=2)$ state
is populated, while in the electric trap there no longer is a
preferred orientation axis and the atoms redistribute over the
degenerate $m_F$ sublevels. However, the dipole component of the
electric field is dominant at the trap center, and therefore the
field points downwards along the $z$ axis across the whole imaging
region. The probe beam is linearly polarized with its polarization
vector almost parallel to this axis. In general, if linearly
polarized light is aligned with the quantization axis, there are
no circular polarization components. Therefore, the probe beam
drives only transitions with $\Delta m_F=0$, as indicated in the
inset in Fig.~\ref{fig:StarkShift}. Hence, no transitions to the
$m_F=\pm 3$ levels of the upper $5\,^{2}P_{3/2}(F=3)$ state are
possible. The strength of the allowed transitions varies as shown
in the inset in Fig.~\ref{fig:StarkShift} and is accounted for in
the simulations by weighting the transitions accordingly. The
natural linewidth of the transition is 6 MHz; therefore the beam
can simultaneously excite transitions with different $m_F$, if the
upper sublevels are close enough in frequency. The linewidth of
the laser is in the submegahertz regime and thus has a negligible
contribution.

From the simulations we determine that the actual fields in the
$\rho$-focusing configuration are lower than expected for the
ideal geometry presented in Fig.~\ref{fig:PictureTrap}. In
contrast, the simulations for $z$ focusing match well. These lower
electric fields can be reproduced by moving the lower end cap
electrode down by 0.25 mm in the simulations. The calculations
with the retracted end cap electrode are shown in
Fig.~\ref{fig:InTrap} and they agree remarkably well with the
corresponding experimental images for both trapping
configurations. The displacement of the end cap results in
considerable changes only for the $\rho$-focusing fields. In the
$z$-focusing configuration, it is the ring electrodes that are
primarily important for creating the maximum along $z$, whereas
the end cap voltages are relatively low.

By comparing the simulations with the measurements, we conclude
that in-trap imaging of the atoms is a very convenient method to
monitor the electric fields in the trap. Using the simulations we
are able to verify the geometry of our trap and determine possible
inaccuracies. All further simulations discussed in the paper are
carried out using the fields associated with the retracted end cap
geometry.

\section{Formation of the trapped cloud}\label{sec:CloudEvolution}

\begin{figure}[pt]
\includegraphics[scale=0.4]{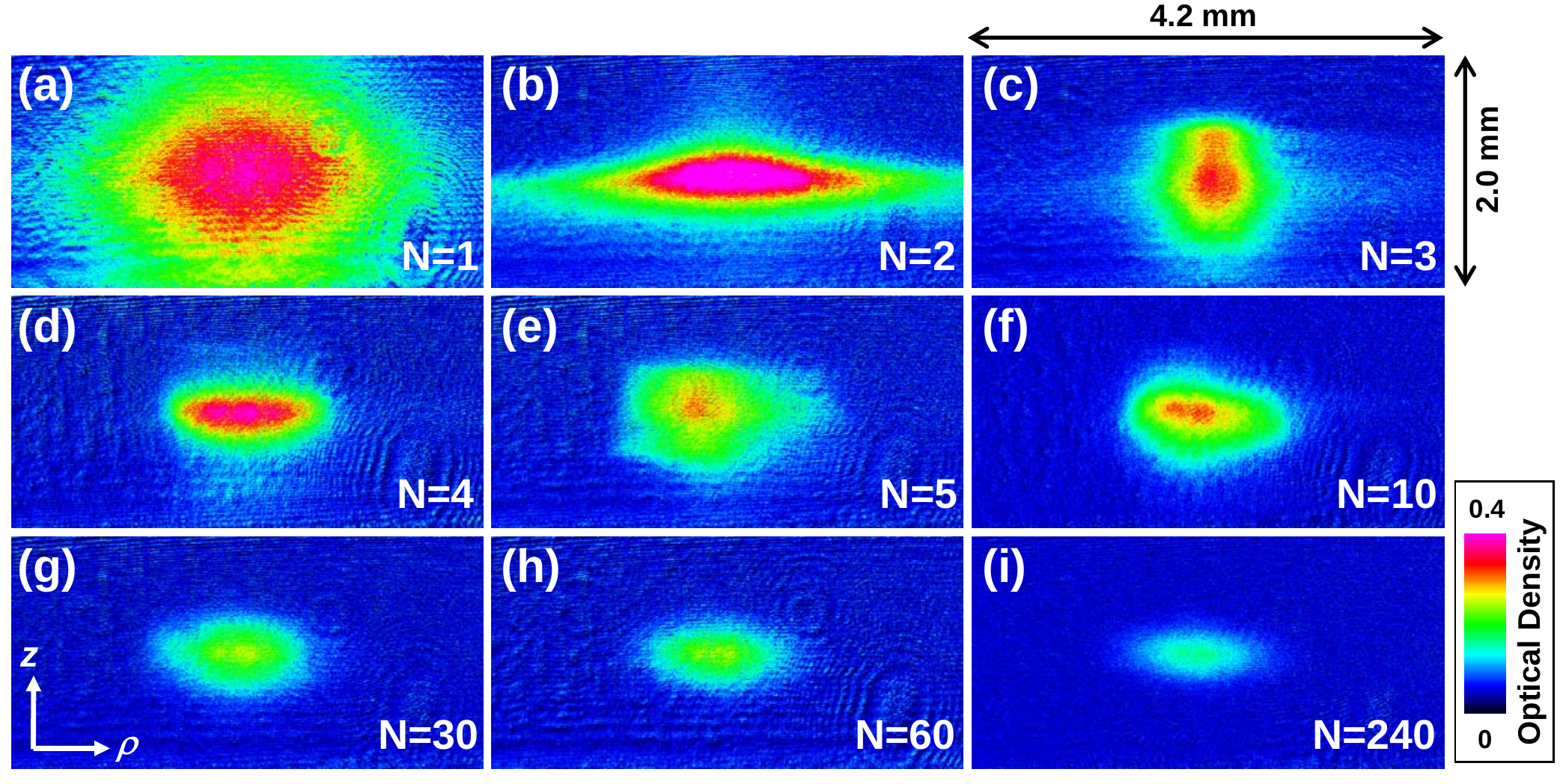}
\caption{Gradual formation of the trapped atom cloud. The pictures
are taken at a switching frequency of 60 Hz and show the cloud
after $N =$ 1, 2, 3, 4, 5, 10, 30, 60, and 240 full switching
cycles. All images are averaged five times. The corresponding
optical density is indicated by the color scale.}
\label{fig:Figure1}
\end{figure}

We will now examine the onset of stable ac electric trapping by
imaging cloud shapes after a small number of switching cycles.
Figure \ref{fig:Figure1} shows the gradual formation of a trapped
cloud as an increasing number $N$ of full switching cycles is
applied. The absorption images always show the 2 mm gap between
the ring electrodes. In Fig.~\ref{fig:Figure1}(a) only one
switching cycle at 60 Hz has been employed and the atoms
completely fill the space available to them. Note that without
application of this switching cycle the atoms would have by now
fallen out of view due to gravity. After $N=2$ switching cycles,
as displayed in Fig.~\ref{fig:Figure1}(b), the shape of the cloud
has changed dramatically. Compared to Fig.~\ref{fig:Figure1}(a),
the number of atoms is now much smaller and the cloud is almost
pancake shaped. In Fig.~\ref{fig:Figure1}(c) after $N=3$ switching
cycles, the cloud has taken a rounder shape. The 'final' shape
emerges only after $N=4$ switching cycles, as can be seen in
Figs.~\ref{fig:Figure1}(d) and \ref{fig:Figure1}(e). However, most
of the atoms are still metastably trapped and will finally escape
from the trap. After $N=5$ and $N=10$ switching cycles, the cloud
exhibits an asymmetric shape, which is visible in
Figs.~\ref{fig:Figure1}(e) and \ref{fig:Figure1}(f), respectively.
This asymmetry can be attributed to possible misalignments between
the center of the magnetic trap and the center of the ac trap.
Even after $N=30$ switching cycles [Fig.~\ref{fig:Figure1}(g)] the
shape is not as smooth as in Fig.~\ref{fig:Figure1}(h) after
$N=60$ switching cycles. A small feature is still visible on the
left side of the cloud. However, the shape remains unchanged at
longer trapping times, as illustrated in Fig.~\ref{fig:Figure1}(i)
where the number of atoms has decreased due to collisions with the
background gas.

\begin{figure}[pt]
\includegraphics[scale=0.5]{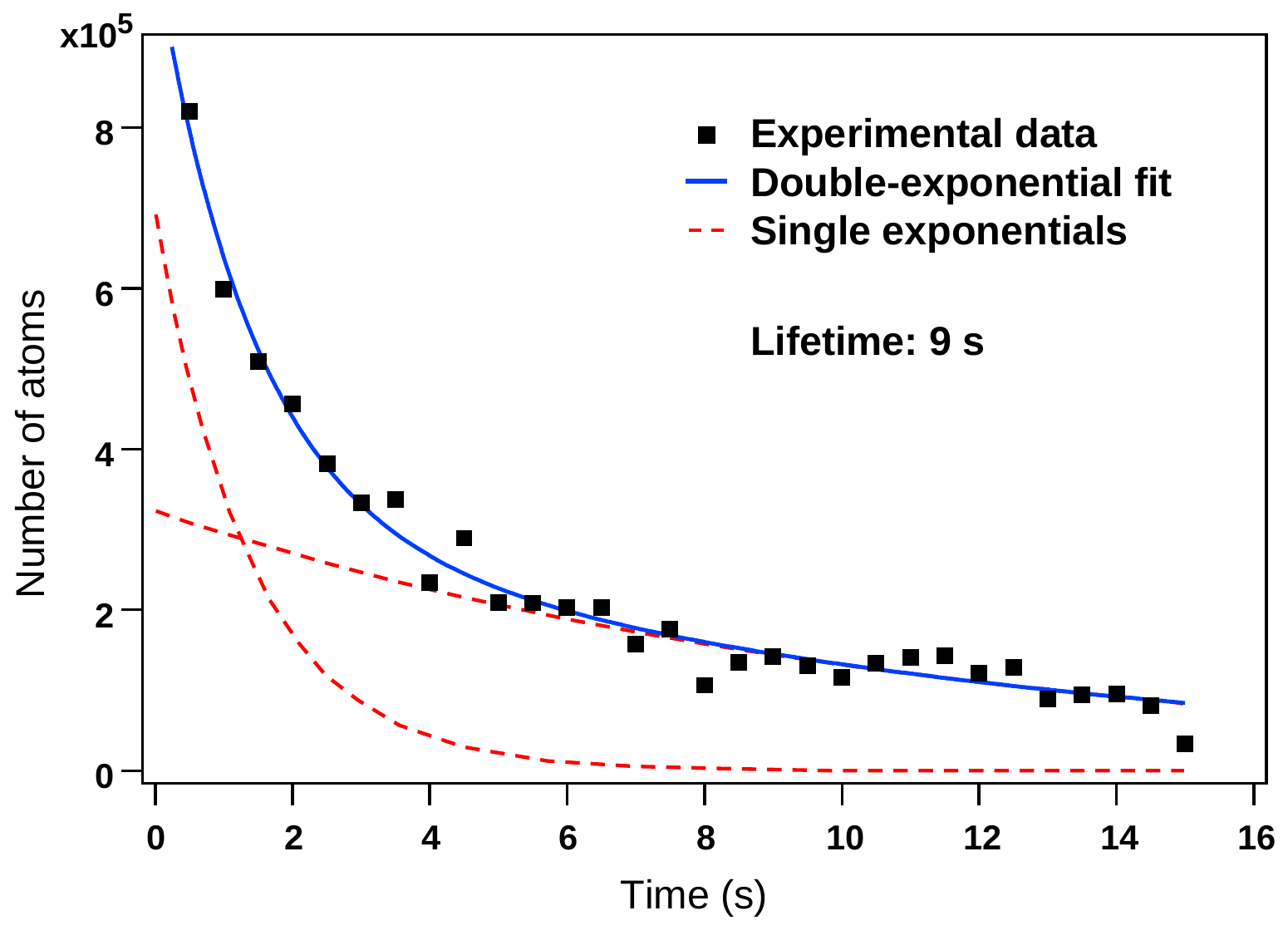}
\caption{Lifetime measurement of the ac trapped atom cloud. The
experiment is performed at 60 Hz, and all the data points are
single-shot measurements. The solid line is a double-exponential
fit to the data points, while the dashed curves show each
exponential individually.}
\label{fig:Lifetime}
\end{figure}

We also studied the atom loss from the trap more quantitatively by
recording the number of atoms in the first 0.5~s of trapping. The
fast decrease during the first ten switching cycles observed in
the images can be fitted by an exponential yielding a $1/e$
lifetime of about 100 ms. This can be attributed to the fact that
most of the metastably trapped atoms leave the trap region within
the first few switching cycles.

Figure \ref{fig:Lifetime} illustrates the lifetime in the ac trap,
which is limited by the collisions with the background gas.
Displayed is the number of atoms versus the trapping time in
seconds. We image the atom cloud at trapping times of up to 15 s,
which corresponds to $N=900$ switching cycles at the switching
frequency of 60 Hz. The data points are fit using a double
exponential yielding a value of 9 s for the lifetime and a value
of 1 s for the fast decay. Similarly to the 100 ms decay in the
previous paragraph, this fast decay accounts for metastably
trapped atoms. These atoms, however, survive in the trap much
longer, as they almost have the correct initial conditions to be
stably trapped. The 9 s lifetime value is consistent with
measurements of a magnetically trapped cloud, performed in the
same vacuum chamber.

We therefore conclude that at least 60 switching cycles have to be
used in the experiment. Otherwise, the dynamics would be mainly
guided by the behavior of metastably trapped atoms. Imaging the
atoms after a trapping time of 1~s is a good compromise between
stable behavior of the atoms and good signals, as longer trapping
times suffer from atom loss due to background collisions.

\section{Trap performance}\label{sec:Performance}

In this section, we first visualize the dynamic confinement of the
atoms in the trap by looking at the atomic distribution at
different times within a switching cycle. From this we can
qualitatively understand the motion in the trap. Next, we study
the dependence of the number of trapped atoms on the switching
frequency. Finally, we analyze the asymmetry of the switching
cycle and present measurements on the mean kinetic energy of the
atoms.

\begin{figure}[pt]
\includegraphics[scale=0.37]{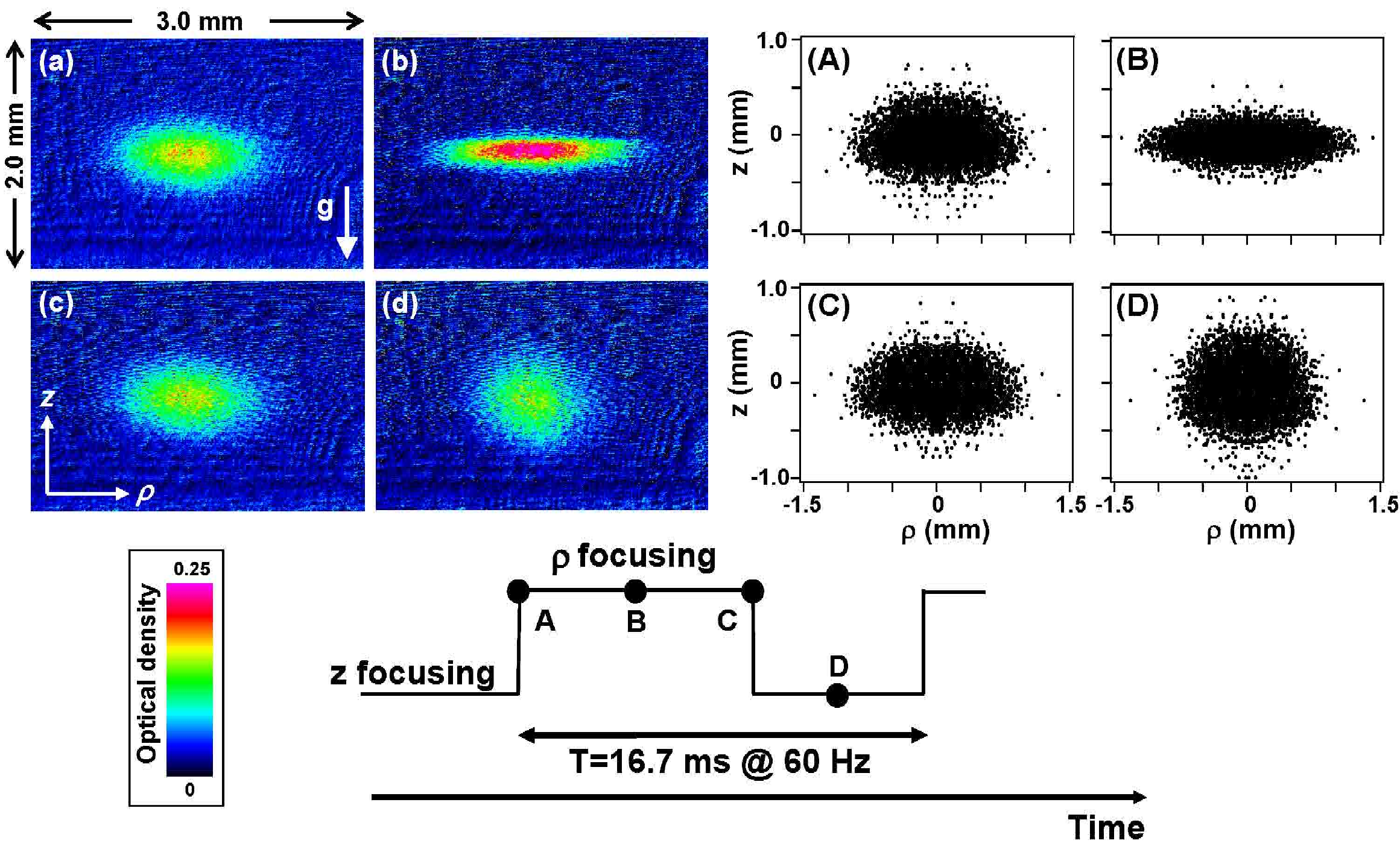}
\caption{Absorption images (a)$-$(d) of the atom cloud at
different phases within the 61st switching cycle at a trapping
frequency of 60 Hz. The pictures (A)$-$(D) are simulation results
of the atomic distribution, labelled with matching capital
letters. The schematic shows the applied switching cycle, where
the appropriate phases are also indicated. The measurements are
averaged five times, and the optical density is shown in the color
scale.} \label{fig:TrapPhases}
\end{figure}

Characteristic for an ac trap is the fact that trapping is
dynamic, i.e., the atoms are forced to move during the switching
cycle, which is referred to as micromotion in ion traps. This can
be seen in Fig.~\ref{fig:TrapPhases} where the images on the left
show the atom cloud at different phases within the 61st switching
cycle for a trapping frequency of 60~Hz. The corresponding phases
A$-$D are indicated on the switching cycle, which is asymmetric
with 59\% of $\rho$ focusing followed by 41\% of $z$ focusing.

In Fig.~\ref{fig:TrapPhases}(a) the atoms have just experienced
$z$ focusing and they are therefore moving inwards along $z$ and
outwards along $\rho$. Consequently, in the middle of $\rho$
focusing, the cloud is focused in $z$ and elongated in $\rho$, as
shown in Fig.~\ref{fig:TrapPhases}(b). However, the
$\rho$-focusing forces have decelerated the motion along both axes
and the atoms are now at the turning point of the micromotion
before they change direction. Due to this motion, inwards in
$\rho$ and outwards in $z$, the cloud in
Fig.~\ref{fig:TrapPhases}(c) has a shape similar to that in
Fig.~\ref{fig:TrapPhases}(a). But as the atoms have just
experienced the $\rho$-focusing phase, the velocity components are
now pointing towards the center of the trap in $\rho$ and outwards
in $z$. This leads to a contraction in $\rho$ as shown in
Fig.~\ref{fig:TrapPhases}(d), which is recorded in the middle of
$z$ focusing where the cloud shape is round. As in
Fig.~\ref{fig:TrapPhases}(b), the atoms are at a standstill before
they turn around. At the end of the switching cycle the cloud
shape will be identical to the one in
Fig.~\ref{fig:TrapPhases}(a). Figures
\ref{fig:TrapPhases}(A)$-$\ref{fig:TrapPhases}(D) show simulations
of the atomic distribution at the same switching times within the
61st cycle as the experimental data
Figs.~\ref{fig:TrapPhases}(a)$-$\ref{fig:TrapPhases}(d). The
agreement between the simulations and the experimental data is
very good.

\begin{figure}[pt]
\includegraphics[scale=0.53]{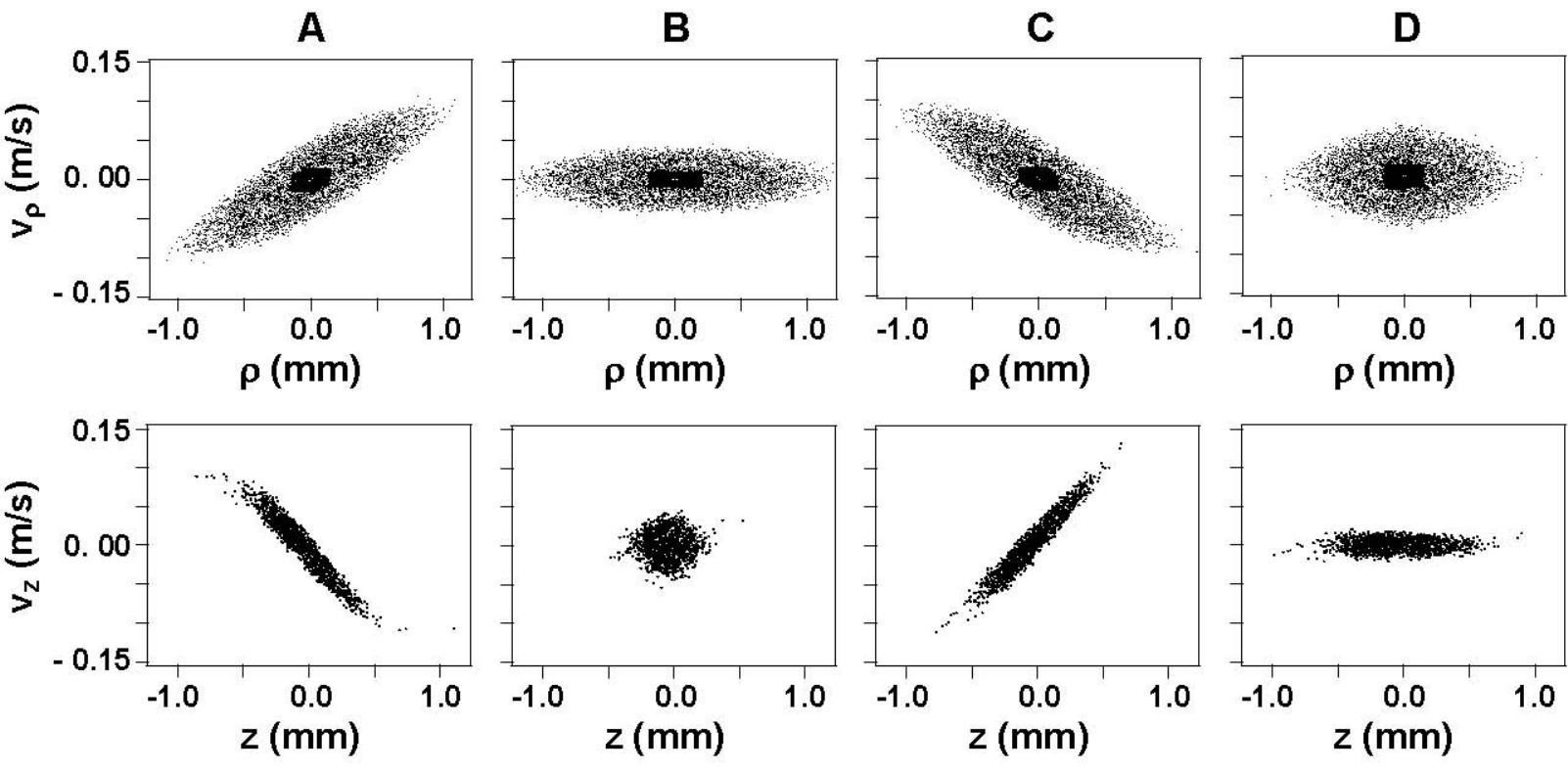}
\caption{Radial (first row) and axial (second row) simulated
phase-space distributions of the atoms at different trap phases
during the 61st switching cycle. The phase space plots are
obtained for the same switch-off times A$-$D used in
Fig.~\ref{fig:TrapPhases}, as indicated for each column. Note that
for the simulations in $\rho$ a finer grid is used for small
velocities and small positions, thus increasing the density of
particles around zero.} \label{fig:PhaseSpace}
\end{figure}

Figure \ref{fig:PhaseSpace} illustrates the evolution of the
phase-space distributions during the 61st switching cycle. The
radial velocity $v_{\rho}$ is plotted versus $\rho$ in the first
row, and the axial velocity $v_z$ is plotted versus $z$ in the
second row. These plots result from the same simulations carried
out for the position distributions in
Figs.~\ref{fig:TrapPhases}(A)$-$\ref{fig:TrapPhases}(D) and show
the corresponding radial and axial phase-space distributions. The
capital letters indicate for each column the appropriate
switch-off time. For the simulations in $\rho$, a finer grid is
used for the values around zero in order to eliminate possible
numerical inaccuracies.

For both coordinates $\rho$ and $z$, the distribution is
oscillating in both position and velocity. For the pictures in the
first row, the distribution is rotating clockwise from A via B to
C; for the second row the same happens from C via D to A. The
velocity spread is maximal for A and C where the distribution is
tilted by 45$^{\circ}$. Some particles are moving at a speed of
more than 0.1 m/s, which corresponds to a temperature of about 50
$\mu$K. Note that the spread in $\rho$ and $v_{\rho}$ is larger
than the spread in $z$ and $v_z$. As mentioned before, the atoms
are at the turning point of the micromotion when they are in the
middle of the focusing (defocusing) stage. The spread in position
is maximal in the middle of the focusing stage, i.e., along $\rho$
in B and along $z$ in D, as also seen in the corresponding
pictures in Fig.~\ref{fig:TrapPhases}. On the other hand, the
velocity spread is always minimal at these points, i.e., for
$v_{\rho}$ in B and for $v_z$ in D. In the middle of defocusing,
the cloud shape becomes round, as seen for $\rho$ in D and for $z$
in B. Note that the phase-space distributions shown are equivalent
to the phase-space acceptance of the trap at that particular
moment in the switching cycle. These phase-space distributions
have been analyzed in detail for the case of molecules moving in
the same trapping configuration \cite{Jacqueline06b}.

\begin{figure}[pt]
\includegraphics[scale=0.44]{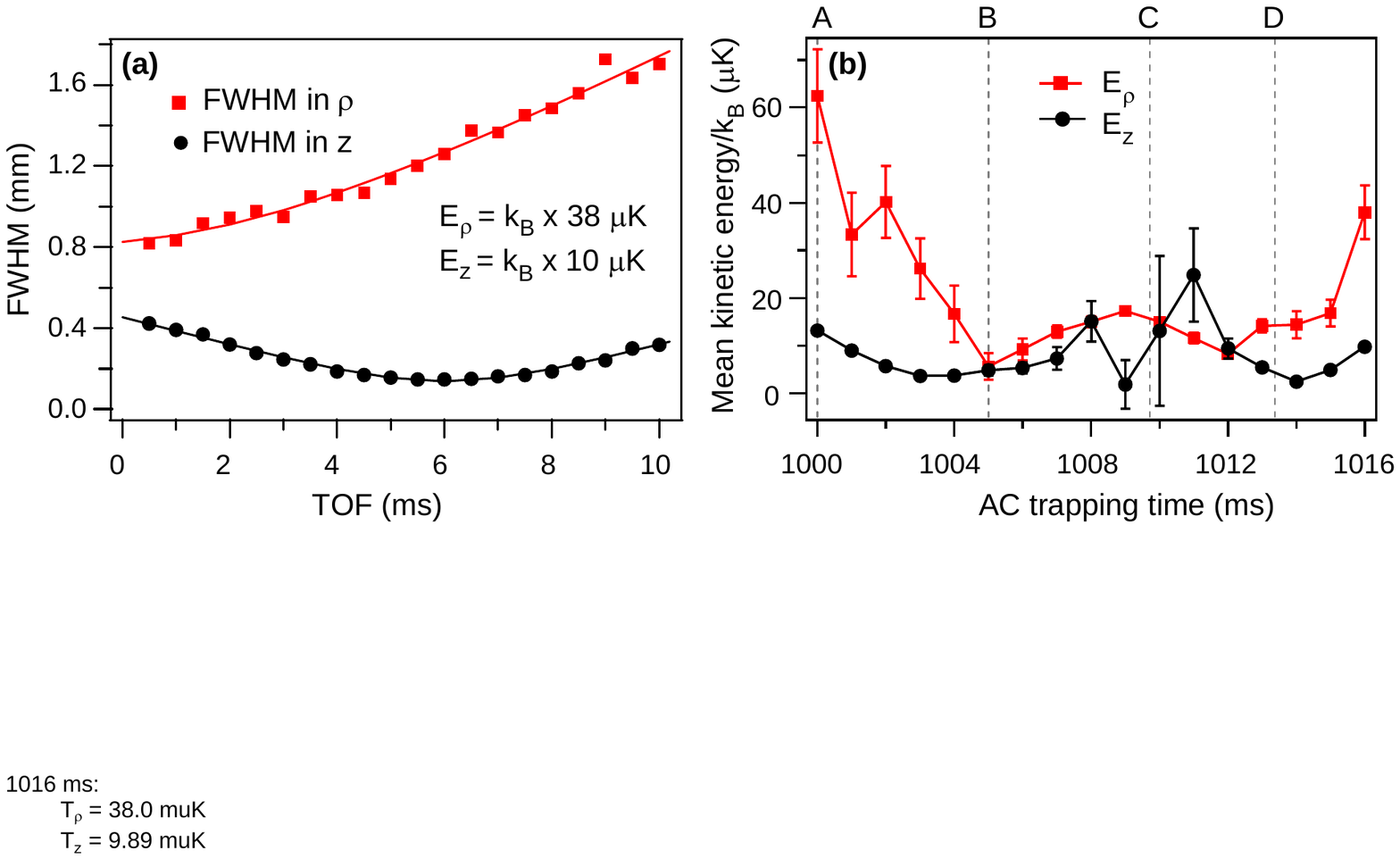}
\caption{(a) Ballistic expansion measurements of the ac trapped
cloud after a trapping time of 1016 ms at 60 Hz. Plotted are the
measured FWHM of the cloud in the $\rho$ (squares) and $z$
(circles) directions versus the TOF. The solid lines show the
corresponding fits. (b) Mean kinetic energies $E_\rho$ (squares)
and $E_z$ (circles) of the atoms for various ac trapping times
within the 61st switching cycle. The dashed lines indicate the
times A$-$D in the switching cycle. All data points are plotted
together with the associated error bars, which are the standard
deviations of the fit.}
\label{fig:TOF}
\end{figure}

To characterize the mean kinetic energy of the cloud, we perform
ballistic expansion measurements of the trapped atoms. These
time-of-flight (TOF) measurements are carried out at various trap
phases within the 61st switching cycle at our standard switching
frequency of 60 Hz. Figure \ref{fig:TOF}(a) shows a typical TOF
series where the cloud is imaged after a trapping time of 1016 ms,
i.e., towards the end of the $z$-focusing phase. Plotted is the
measured full width at half maximum (FWHM) of the cloud versus the
TOF. The kinetic energy is determined from a fit to the measured
FWHM, where the initial density and velocity distributions are
assumed to be Gaussian. The fit yields a value of $E_\rho = k_{B}
\times 38$~$\mu$K in the radial direction (squares), and a value
of $E_z = k_{B} \times 10$~$\mu$K in the axial direction (circles)
where $k_{B}$ is the Boltzmann constant. Note that the size of the
cloud in the axial direction first decreases to a minimum value at
5 ms, because the atoms are moving towards the center at this
particular switch-off time. Then, the cloud FWHM increases as the
atoms continue to move along $z$ and the cloud spreads out, i.e.,
atoms originally at the top of the cloud are now at the bottom,
and vice versa.

Figure \ref{fig:TOF}(b) shows the mean kinetic energy in the
radial and the axial directions, $E_\rho$ and $E_z$, for various
ac trapping times within the same 61st switching cycle. The error
bars indicate the standard deviations of the fit. The dashed lines
correspond to the phases A$-$D of the switching cycle. As expected
from Fig.~\ref{fig:PhaseSpace}, the velocities in the radial
direction (squares) are higher than the velocities in the axial
direction (circles) which is confirmed by our measurements. We
also expect that the kinetic energy is minimal in the middle of
the focusing phases, i.e., at the turning points of the
micromotion, B and D. Only at these phases of the switching cycle
can we exclude the contribution of the micromotion to the kinetic
energy. From measurements at point B and close to point D, we
obtain minimum values of about $k_{B} \times 10$~$\mu$K for the
mean kinetic energy.

As the atoms do not have well-defined Gaussian velocity
distributions across the entire switching cycle, the FWHM fits of
the expanding cloud can have large error bars. This is primarily
the case for ac trapping times where the motion of the atoms is
governed by the defocusing forces, i.e., from the middle of the
defocusing phase to the middle of the next focusing phase.
Therefore, as seen in Fig.~\ref{fig:TOF}(b), in the radial
direction we obtain larger error bars  for trapping times between
D and B via A. Here we take into account the fact that the
switching cycle wraps around, i.e., the end of one cycle coincides
with the beginning of the next cycle. For the axial direction, the
error bars are largest for trapping times between B and D, that is
to say for the complementary part of the switching cycle.

\begin{figure}[pt]
\includegraphics[scale=0.45]{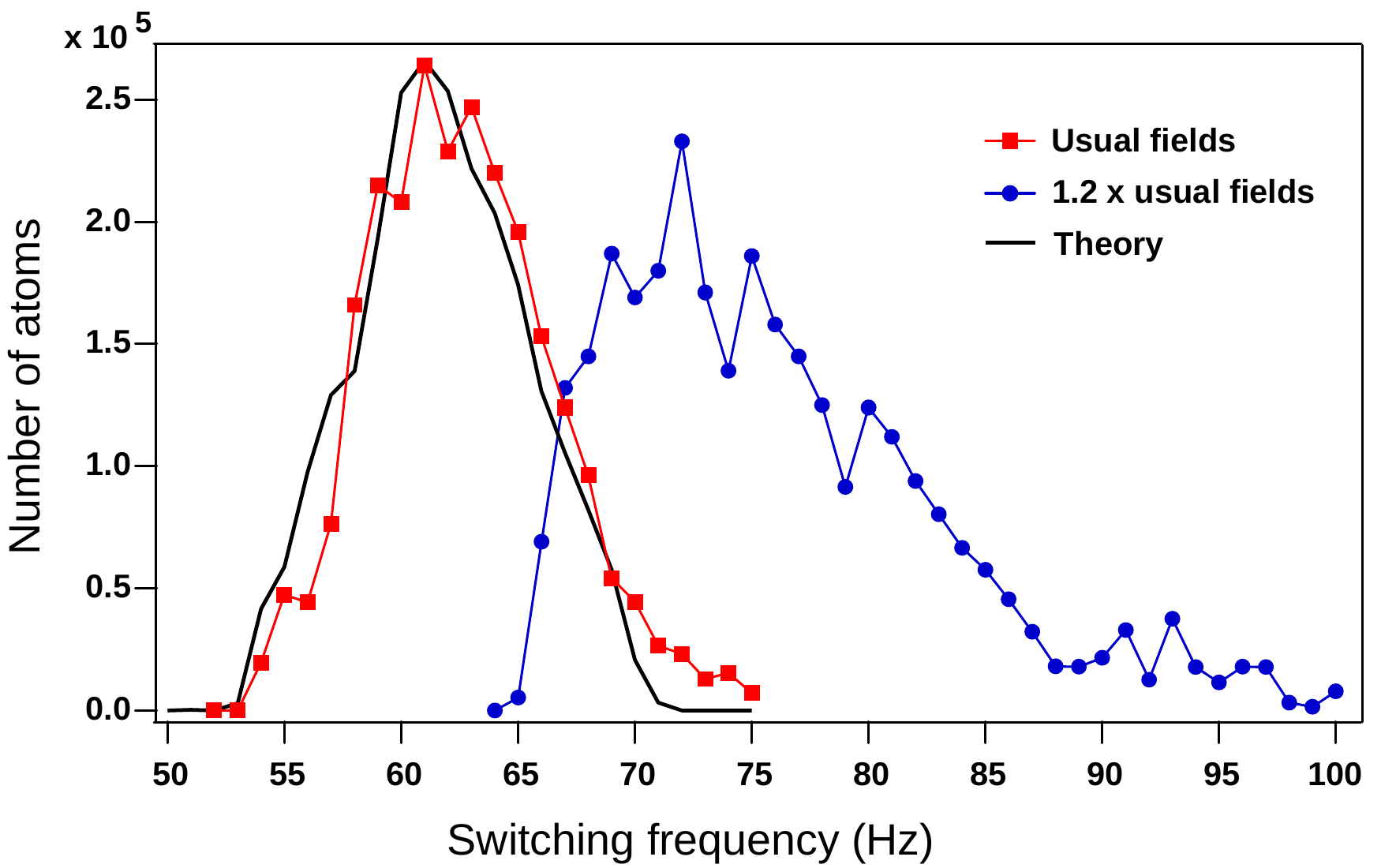}
\caption{Frequency scan for two different sets of voltages. The
squares show measurements where the voltages from
Fig.~\ref{fig:PictureTrap} are employed. The solid curve is the
corresponding theory prediction. The circles display measurements
where higher voltages are used, resulting in electric fields that
are 1.2 times stronger. The frequencies for stable trapping are
then shifted to higher values. The number of atoms is measured
after a trapping time of 5 s, and ten pictures are averaged for
each data point.} \label{fig:FrequencyScan}
\end{figure}

Figure \ref{fig:FrequencyScan} shows the number of trapped atoms
versus the applied switching frequency after a 5 s trapping time.
Two different sets of voltages are used. The frequency scan
plotted with squares was obtained using the voltages indicated in
Fig.~\ref{fig:PictureTrap}. For the measurement plotted with
circles, the applied voltage set results in an electric field that
is a factor of 1.2 higher. The solid curve is a simulation carried
out using the lower-voltage set which is in excellent agreement
with the experimental data. For the lower voltages, trapping works
in a rather narrow range between 54 and 75 Hz with a maximum of
$2.5 \times 10^5$ atoms trapped at 61 Hz after 5 s. No trapping is
observed below the 54 Hz threshold. For the higher voltages,
trapping occurs at higher switching frequencies, as expected. As
with the lower voltages, there is a strong increase in signal
above the threshold frequency of 64 Hz, and a slowly decreasing
signal for the higher frequencies. Here, the range of working
frequencies is broader. Despite the deeper trapping potential at
higher voltages, the recorded number of trapped atoms is smaller,
which we attribute to tiny discharges that we did not observe for
the lower voltages. These discharges lead to a local increase in
pressure which reduces the number of trapped atoms due to a higher
rate of background collisions.

\begin{figure}[pt]
\includegraphics[scale=0.58]{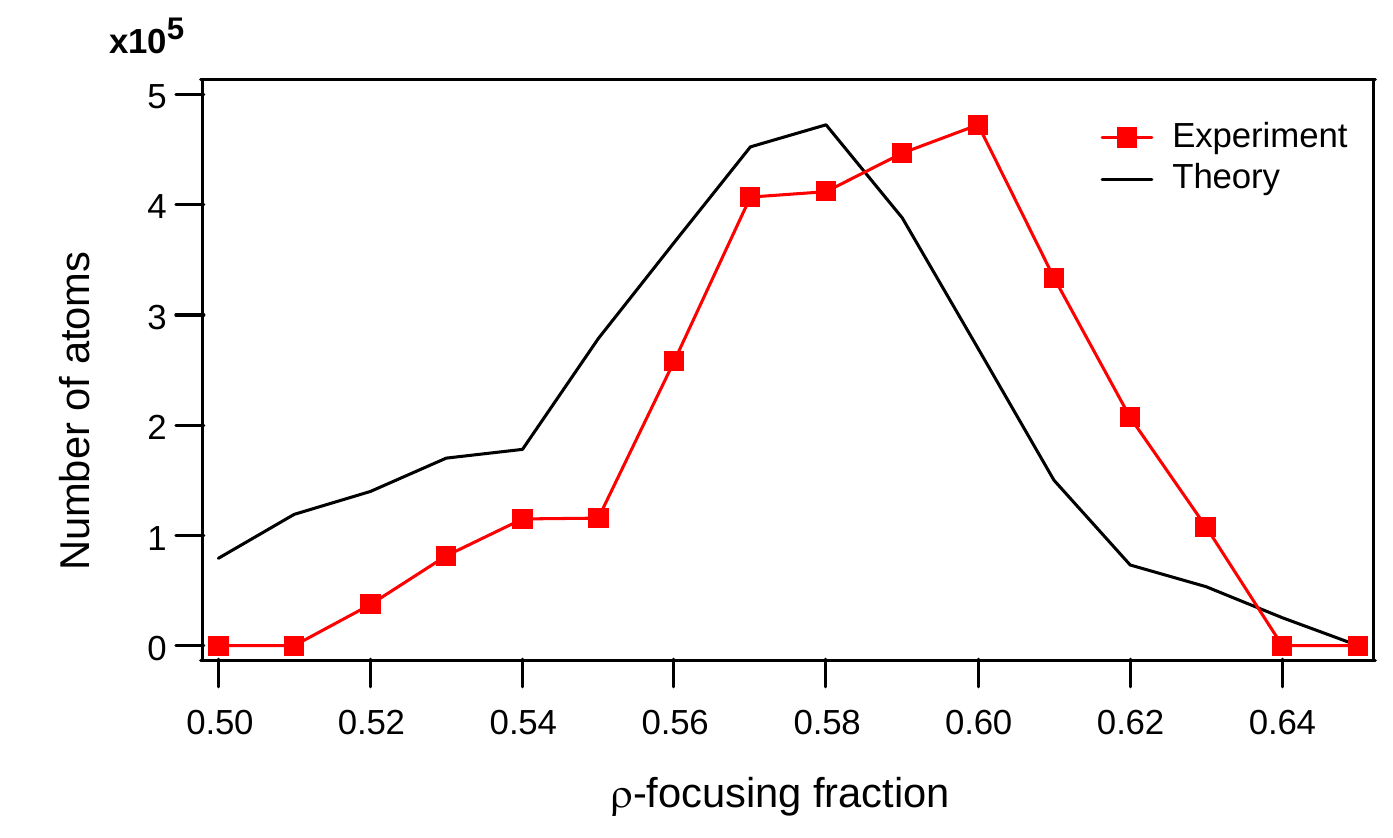}
\caption{Atom number versus $\rho$-focusing fraction in a
switching cycle of 16.7 ms corresponding to a switching frequency
of 60 Hz. The data are taken after 1 s of trapping, and the
measurements are averaged five times. The solid black curve is the
simulation result.}
\label{fig:PercentageOn}
\end{figure}

In Fig.~\ref{fig:PercentageOn} the relative switching time of
$\rho$ versus $z$ focusing is varied, while the duration of the
switching cycle is held constant at 16.7 ms. The first data point
refers to a symmetric switching cycle with 50\% of $\rho$ focusing
and 50\% of $z$ focusing, where no signal is observed. The number
of trapped atoms increases with increasing $\rho$-focusing time.
The maximum atom number after 1~s of trapping is $4.5 \times 10^5$
and is found for a switching cycle with 60\% of $\rho$ focusing.
For longer $\rho$-focusing times the number of atoms quickly
decreases with no signal observed from 64\% onwards. The solid
curve is the corresponding theory prediction and shows the same
trend as the experimental data. There is a clear shift, however,
between theory and experiment, with the theoretical maximum at a
smaller value of the $\rho$-focusing fraction. The asymmetry of
the switching cycle is partly explained by the need to compensate
for differences between the ideal trapping geometry and the actual
experimental configuration. The experimental misalignments are
likely more complex than the already mentioned retracted end cap
geometry, which is taken into account in our simulations.

\section{Probing the phase-space acceptance of the trap}\label{sec:PhaseJump}

The phase-space acceptance can be probed by introducing a sudden
change in the switching cycle. Most of the atoms have the wrong
initial conditions to survive this change and they will be lost
from the trap. Only the few atoms that reside in the accepted part
of the phase-space distribution will remain trapped if more
switching cycles are applied afterwards.

\begin{figure}[pt]
\includegraphics[scale=0.48]{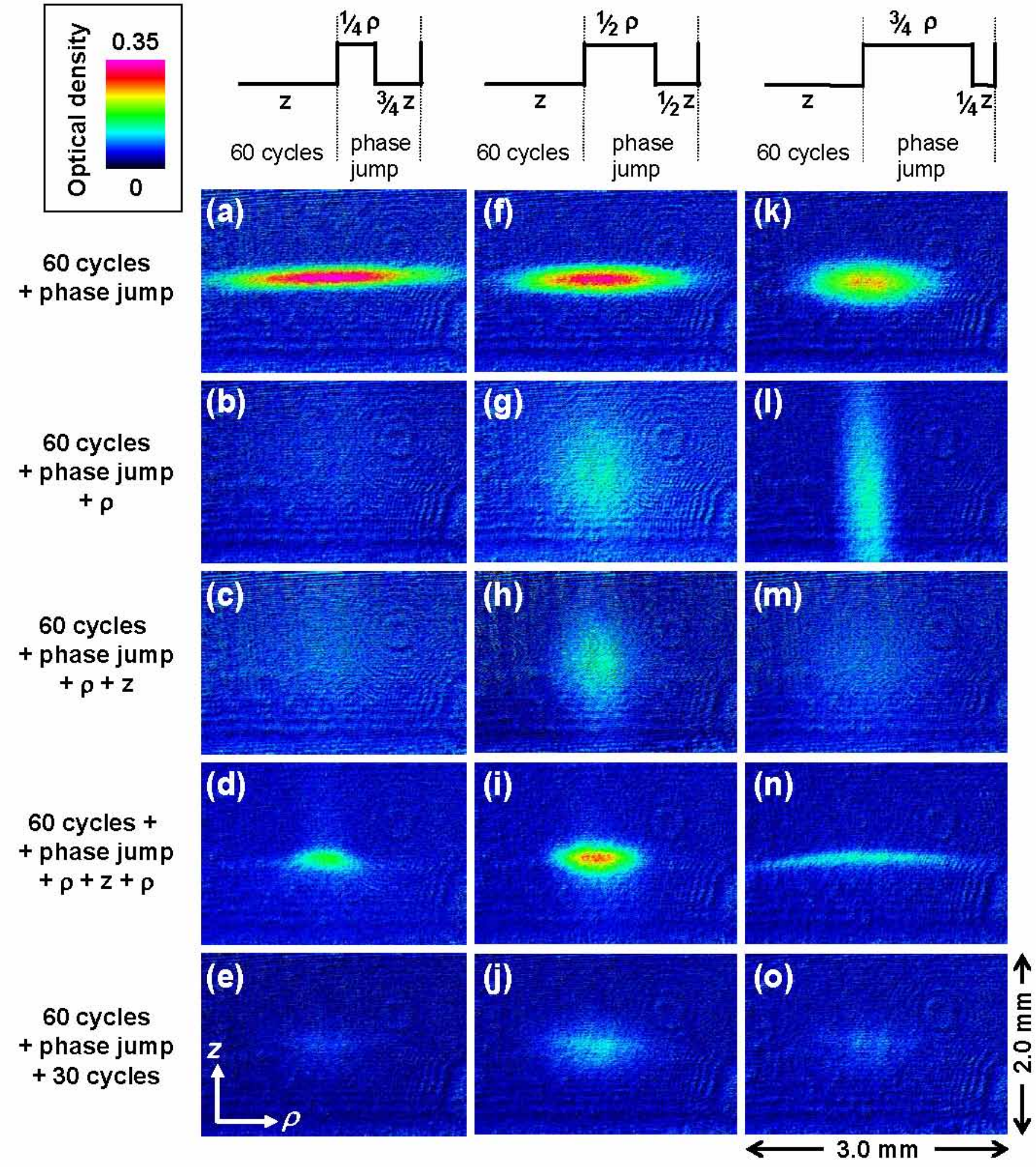}
\caption{Pictures of the atom cloud after 60 switching cycles at
60 Hz followed by a phase jump. For the first column (a)$-$(e),
$\rho$ focusing is on for only $1/4$ of its usual duration and $z$
focusing is on for $3/4$ of its usual setting. For the second
column (f)$-$(j), a symmetric phase jump is performed with $1/2$
$\rho$ focusing and $1/2$ $z$ focusing. In the third column
(k)$-$(o), $3/4$ of $\rho$ focusing is followed by $1/4$ of $z$
focusing. Each column shows the atoms imaged directly after the
phase jump (first row), after an additional $\rho$-focusing phase
applied after the phase jump (second row), after a full switching
cycle applied after the phase jump (third row), after a full cycle
and a $\rho$-focusing phase (fourth row), and when 30 full
switching cycles have been applied after the phase jump (last
row). All pictures are averaged five times.} \label{fig:PhaseJump}
\end{figure}

For this experiment, a truncated switching cycle is applied to the
stably trapped atom cloud. We will refer to this modified
switching cycle as a 'phase jump'. First, 60 switching cycles with
the usual 59\% of $\rho$ focusing and switching frequency of 60 Hz
are applied. This ensures that most of the metastably trapped
atoms have escaped from the trap. Then comes a cycle with reduced
durations for both the $\rho$- and the $z$-focusing phases. This
is illustrated in Fig.~\ref{fig:PhaseJump}. In the first column, a
phase jump is applied where the $\rho$ focusing is reduced to
$1/4$ and the $z$ focusing is reduced to $3/4$ of their usual
durations. For the measurements in the second column, a symmetric
phase jump is applied, i.e., a $1/2$ $\rho$-focusing phase is
followed by a $1/2$ $z$-focusing phase. In the third column, the
measurements are again taken for an asymmetric phase jump with
$3/4$ of $\rho$ focusing and $1/4$ of $z$ focusing.

In the first row of Fig.~\ref{fig:PhaseJump}, the cloud is imaged
directly after the phase jump. From left to right, the radial
extent of the cloud decreases while the cloud becomes larger in
$z$. In Fig.~\ref{fig:PhaseJump}(a) the atoms have been exposed to
$z$ focusing longer than to $\rho$ focusing, which keeps the cloud
tightly together in $z$, whereas the atoms spread out in $\rho$.
In Fig.~\ref{fig:PhaseJump}(k), due to the short $z$-focusing
phase, the cloud is spread out in $z$ and confined in $\rho$. In
Fig.~\ref{fig:PhaseJump}(f) the phase jump is symmetric and the
cloud shape is intermediary between the situations in
Figs.~\ref{fig:PhaseJump}(a) and \ref{fig:PhaseJump}(k). The
pictures in the second row show the atoms after the application of
an additional, full $\rho$-focusing phase. As a consequence, from
Fig.~\ref{fig:PhaseJump}(b) via Fig.~\ref{fig:PhaseJump}(g) to
Fig.~\ref{fig:PhaseJump}(l), the atoms are more focused in the
$\rho$ direction. In Fig.~\ref{fig:PhaseJump}(b) the cloud density
is very low making the atoms barely visible. In
Fig.~\ref{fig:PhaseJump}(g) the cloud is fairly well confined
because the atoms are not as perturbed as in
Figs.~\ref{fig:PhaseJump}(b) and \ref{fig:PhaseJump}(l) due to the
symmetric switching. In Fig.~\ref{fig:PhaseJump}(l) the cloud is
spreading out along $z$ as it has experienced only a short
$z$-focusing phase. In the third row a subsequent, full
$z$-focusing stage is applied so that the atoms have by now
experienced one full switching cycle after the phase jump. In
Fig.~\ref{fig:PhaseJump}(c) the density has increased due to the
$z$ focusing, and a very dilute cloud is now visible. In
Fig.~\ref{fig:PhaseJump}(h) the cloud is pretty well confined. In
Fig.~\ref{fig:PhaseJump}(m) atoms have escaped from the trap along
the $z$ direction, the additional defocusing in $\rho$ during the
$z$-focusing phase leading to a dilute cloud. In the fourth row an
additional $\rho$-focusing phase is applied. For both asymmetric
phase jumps the atoms now form a small cloud. Very interesting is
the pancake shape in Fig.~\ref{fig:PhaseJump}(n), elongated along
$\rho$ and focused in $z$ due to the prior phase of $z$ focusing.
Looking at the series in
Figs.~\ref{fig:PhaseJump}(l)$-$\ref{fig:PhaseJump}(n), we notice
that the cloud changes dramatically, being $\rho$ focused in
Fig.~\ref{fig:PhaseJump}(l) and dilute in
Fig.~\ref{fig:PhaseJump}(m) because of the reshaping. The last row
shows the atoms after 30 full switching cycles have followed the
phase jump. For the asymmetric switching procedures in
Figs.~\ref{fig:PhaseJump}(e) and \ref{fig:PhaseJump}(o) few atoms
survive, whereas in the case of symmetric switching more atoms
remain confined in the trap.

\begin{figure}[pt]
\includegraphics[scale=0.67]{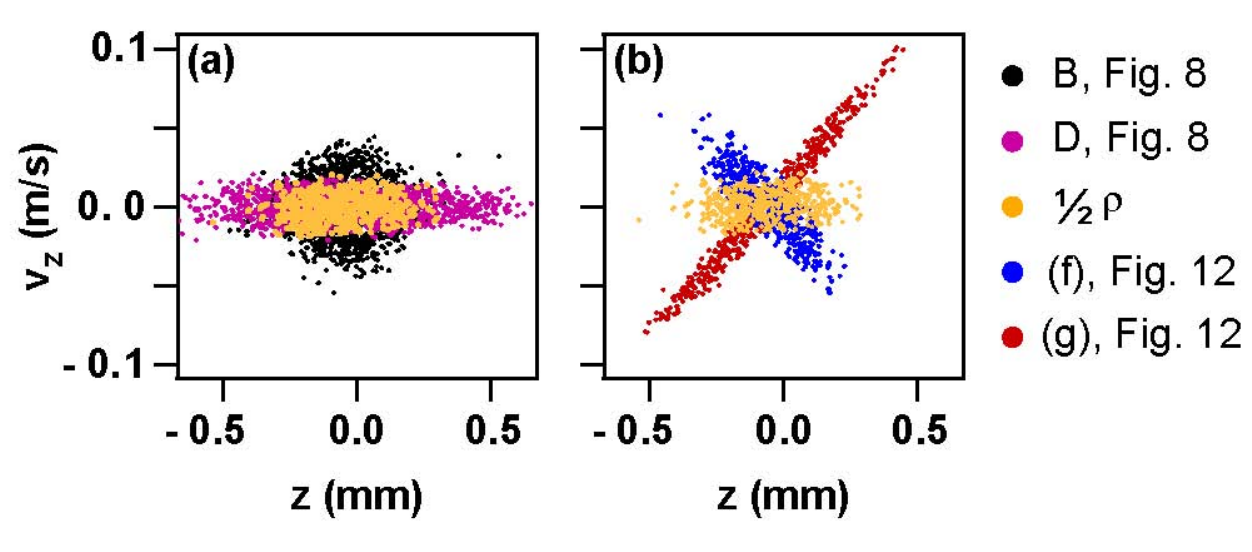}
\caption{Simulated phase-space distributions where $v_z$ is
plotted versus $z$. The figures show those atoms that survive the
symmetric phase jump and the subsequently applied 30 full
switching cycles. In (a), the phase-space distribution after the
$\rho$-focusing stage of the phase jump (orange) is plotted, along
with the distributions already shown in plots B (black) and D
(magenta) of Fig.~\ref{fig:PhaseSpace}. In (b), this distribution
(orange) is shown again, along with the distributions after the
entire phase jump [blue, corresponding to
Fig.~\ref{fig:PhaseJump}(f)], and after an additional
$\rho$-focusing stage [red, corresponding to
Fig.~\ref{fig:PhaseJump}(g)].}
\label{fig:PhaseJumpSimul}
\end{figure}

As discussed for Fig.~\ref{fig:PhaseSpace}, the trap acceptance
varies with the phase in the switching cycle. To visualize how the
motion of the trapped atoms is perturbed by the symmetric phase
jump (i.e., 1/2 $\rho$ focusing followed by 1/2 $z$ focusing), we
carried out simulations which include the phase jump and the
subsequently applied 30 full switching cycles. Note that only the
atoms that survive these final 30 cycles will be discussed in the
following. If this truncated switching cycle is applied, the
distribution and therefore the phase-space acceptance in the
middle of $\rho$ focusing is mapped onto the distribution in the
middle of $z$ focusing. This happens because the second half of
$\rho$ focusing and the first half of $z$ focusing are missing in
the modified cycle. Figure \ref{fig:PhaseJumpSimul} shows the
calculated phase-space distributions, where $v_z$ is displayed
versus $z$. The phase-space distribution after the $\rho$-focusing
stage of the phase jump is presented in
Fig.~\ref{fig:PhaseJumpSimul}(a), along with the distributions
already shown in plots B and in D of Fig.~\ref{fig:PhaseSpace}.
Only those atoms that are in a region of phase space where the
distribution after $1/2$ $\rho$ focusing (black) overlaps with the
distribution after $1/2$ $z$ focusing (magenta) will remain
trapped. Therefore, after the $\rho$-focusing stage of the phase
jump the accepted phase space has reduced to this overlap region
(orange). In Fig.~\ref{fig:PhaseJumpSimul}(b), this reduced
distribution is again illustrated, along with the distributions
after the entire phase jump [blue, corresponding to
Fig.~\ref{fig:PhaseJump}(f)] and after an additional
$\rho$-focusing stage [red, corresponding to
Fig.~\ref{fig:PhaseJump}(g)]. At the end of the phase jump and
therefore at the beginning of the next switching cycle, the
distribution has rotated clockwise by 45$^{\circ}$, as we expect
from plot A in Fig.~\ref{fig:PhaseSpace}. While this distribution
is narrow in both velocity and position, the distribution after an
additional $\rho$-focusing stage starts to spread out thereby
gradually refilling the whole accepted area in phase space.

It becomes clear from these simulations and the associated
measurements that the overlap of the accepted areas in phase space
is optimal for a symmetric phase jump. For asymmetric phase jumps,
more atoms are located in the unaccepted region of phase space and
will therefore be lost from the trap. This is also visible in the
last row of Fig.~\ref{fig:PhaseJump}.

\section{Conclusions}\label{sec:conclusions}

In this paper, we have presented a detailed study on trapping of
rubidium in an ac electric trap. First, we showed that the atoms
can be used to probe the electric fields in the trap, and we
reproduced our results with simulations. By studying the gradual
formation of a trapped cloud, we observed that most of the atoms
that are not stably confined leave the trap within the first
second. In a typical experiment, $3 \times 10^{5}$ atoms are
stably trapped with a lifetime of about 9~s, limited by collisions
with the background gas. One of the nice features of our ac trap
is the ability to directly visualize the atom dynamics at
different phases of the switching cycle using absorption imaging.
Trajectory calculations were carried out to confirm this dynamic
behavior and to understand the corresponding phase-space
distributions. Additionally, the mean kinetic energy of the
trapped cloud was observed to vary across the switching cycle, as
it is dominated by the micromotion in the trap. Values of about 10
$\mu$K were measured at the points where the micromotion does not
contribute. We have also studied the dependence of the trapped
atom number on the switching frequency and the symmetry of the
switching cycle. Stable trapping occurs for a narrow frequency
range around 60~Hz and for an asymmetric switching cycle. Finally,
when a modified switching cycle is applied, the motion of the
atoms in the ac trap can be readily understood from simulations of
the phase-space acceptance at various phases of the switching
cycle.

\begin{acknowledgments}
This work is part of the research program of the ``Stichting voor
Fundamenteel Onderzoek der Materie (FOM)," which is financially
supported by the ``Nederlandse Organisatie voor Wetenschappelijk
Onderzoek (NWO)." A.M.\ would like to thank the Alexander von
Humboldt Foundation for their support. We acknowledge useful
discussions with Sam Meek, Allard Mosk, Achim Peters, and Boris
Sartakov, and technical assistance from Peter Geng.
\end{acknowledgments}

\end{document}